\documentclass[reprint,superscriptaddress]{revtex4-1}
\usepackage{color}
\usepackage{graphicx}
\usepackage{hyperref}
\usepackage{amsmath}
\usepackage{array}
\usepackage{subcaption}
\usepackage[justification=raggedright]{caption}

\begin{document}

\title{Spin-Dependent Weakly-Interacting-Massive-Particle--Nucleon Cross Section Limits from First Data of PandaX-II Experiment}

\date{\today}
\author{Changbo Fu}
\author{Xiangyi Cui}
\affiliation{INPAC and Department of Physics and Astronomy, Shanghai Jiao Tong University, Shanghai Laboratory for Particle Physics and Cosmology, Shanghai 200240, China}
\author{Xiaopeng Zhou}
\affiliation{School of Physics, Peking University, Beijing 100871, China}
\author{Xun Chen}
\affiliation{INPAC and Department of Physics and Astronomy, Shanghai Jiao Tong University, Shanghai Laboratory for Particle Physics and Cosmology, Shanghai 200240, China}
\author{Yunhua Chen}
\affiliation{Yalong River Hydropower Development Company, Ltd., 288 Shuanglin Road, Chengdu 610051, China}
\author{Deqing Fang}
\affiliation{Shanghai Institute of Applied Physics, Chinese Academy of Sciences, 201800, Shanghai, China}
\author{Karl Giboni}
\affiliation{INPAC and Department of Physics and Astronomy, Shanghai Jiao Tong University, Shanghai Laboratory for Particle Physics and Cosmology, Shanghai 200240, China}
\author{Franco Giuliani}
\affiliation{INPAC and Department of Physics and Astronomy, Shanghai Jiao Tong University, Shanghai Laboratory for Particle Physics and Cosmology, Shanghai 200240, China}
\affiliation{Center of High Energy Physics, Peking University, Beijing 100871, China}
\author{Ke Han}
\affiliation{INPAC and Department of Physics and Astronomy, Shanghai Jiao Tong University, Shanghai Laboratory for Particle Physics and Cosmology, Shanghai 200240, China}
\author{Xingtao Huang}
\affiliation{School of Physics and Key Laboratory of Particle Physics and Particle Irradiation (MOE), Shandong University, Jinan 250100, China}
\author{Xiangdong Ji}
\email[Spokesperson: ]{xdji@sjtu.edu.cn}
\affiliation{INPAC and Department of Physics and Astronomy, Shanghai Jiao Tong University, Shanghai Laboratory for Particle Physics and Cosmology, Shanghai 200240, China}
\affiliation{Center of High Energy Physics, Peking University, Beijing 100871, China}
\affiliation{Department of Physics, University of Maryland, College Park, Maryland 20742, USA}
\affiliation{T. D. Lee Institute, Shanghai 200240, China}
\author{Yonglin Ju}
\affiliation{School of Mechanical Engineering, Shanghai Jiao Tong University, Shanghai 200240, China}
\author{Siao Lei}
\author{Shaoli Li}
\affiliation{INPAC and Department of Physics and Astronomy, Shanghai Jiao Tong University, Shanghai Laboratory for Particle Physics and Cosmology, Shanghai 200240, China}
\author{Huaxuan Liu}
\affiliation{School of Mechanical Engineering, Shanghai Jiao Tong University, Shanghai 200240, China}
\author{Jianglai Liu}
\affiliation{INPAC and Department of Physics and Astronomy, Shanghai Jiao Tong University, Shanghai Laboratory for Particle Physics and Cosmology, Shanghai 200240, China}
\author{Yugang Ma}
\affiliation{Shanghai Institute of Applied Physics, Chinese Academy of Sciences, 201800, Shanghai, China}
\author{Yajun Mao}
\affiliation{School of Physics, Peking University, Beijing 100871, China}
\author{Xiangxiang Ren}
\affiliation{INPAC and Department of Physics and Astronomy, Shanghai Jiao Tong University, Shanghai Laboratory for Particle Physics and Cosmology, Shanghai 200240, China}
\author{Andi Tan}
\affiliation{Department of Physics, University of Maryland, College Park, Maryland 20742, USA}
\author{Hongwei Wang}
\affiliation{Shanghai Institute of Applied Physics, Chinese Academy of Sciences, 201800, Shanghai, China}
\author{Jiming Wang}
\affiliation{Yalong River Hydropower Development Company, Ltd., 288 Shuanglin Road, Chengdu 610051, China}
\author{Meng Wang}
\affiliation{School of Physics and Key Laboratory of Particle Physics and Particle Irradiation (MOE), Shandong University, Jinan 250100, China}
\author{Qiuhong Wang}
\affiliation{Shanghai Institute of Applied Physics, Chinese Academy of Sciences, 201800, Shanghai, China}
\author{Siguang Wang}
\affiliation{School of Physics, Peking University, Beijing 100871, China}
\author{Xuming Wang}
\affiliation{INPAC and Department of Physics and Astronomy, Shanghai Jiao Tong University, Shanghai Laboratory for Particle Physics and Cosmology, Shanghai 200240, China}
\author{Zhou Wang}
\affiliation{School of Mechanical Engineering, Shanghai Jiao Tong University, Shanghai 200240, China}
\author{Shiyong Wu}
\affiliation{Yalong River Hydropower Development Company, Ltd., 288 Shuanglin Road, Chengdu 610051, China}
\author{Mengjiao Xiao}
\affiliation{Department of Physics, University of Maryland, College Park, Maryland 20742, USA}
\affiliation{Center of High Energy Physics, Peking University, Beijing 100871, China}
\author{Pengwei Xie}
\affiliation{INPAC and Department of Physics and Astronomy, Shanghai Jiao Tong University, Shanghai Laboratory for Particle Physics and Cosmology, Shanghai 200240, China}
\author{Binbin Yan}
\email[Corresponding author: ]{bbyan@hepg.sdu.edu.cn}
\affiliation{School of Physics and Key Laboratory of Particle Physics and Particle Irradiation (MOE), Shandong University, Jinan 250100, China}
\author{Yong Yang}
\email[Corresponding author: ]{yong.yang@sjtu.edu.cn}
\affiliation{INPAC and Department of Physics and Astronomy, Shanghai Jiao Tong University, Shanghai Laboratory for Particle Physics and Cosmology, Shanghai 200240, China}
\author{Jianfeng Yue}
\affiliation{Yalong River Hydropower Development Company, Ltd., 288 Shuanglin Road, Chengdu 610051, China}
\author{Hongguang Zhang}
\affiliation{INPAC and Department of Physics and Astronomy, Shanghai Jiao Tong University, Shanghai Laboratory for Particle Physics and Cosmology, Shanghai 200240, China}
\author{Tao Zhang}
\author{Li Zhao}
\affiliation{INPAC and Department of Physics and Astronomy, Shanghai Jiao Tong University, Shanghai Laboratory for Particle Physics and Cosmology, Shanghai 200240, China}
\author{Ning Zhou}
\affiliation{INPAC and Department of Physics and Astronomy, Shanghai Jiao Tong University, Shanghai Laboratory for Particle Physics and Cosmology, Shanghai 200240, China}
\affiliation{Department of Physics, Tsinghua University, Beijing 100084, China}

\collaboration{PandaX-II Collaboration}
\begin{abstract}

New constraints are presented on the spin-dependent WIMP-nucleon interaction
from the PandaX-II experiment, using a data set corresponding to a total
exposure of 3.3$\times10^4$ kg-days. Assuming a standard axial-vector
spin-dependent WIMP interaction with $^{129}$Xe and $^{131}$Xe nuclei, 
the most stringent upper limits on WIMP-neutron cross sections for
WIMPs with masses above 10 GeV/c$^{2}$ are set in all dark matter direct detection
experiments. The minimum upper limit of $4.1\times 10^{-41}$ cm$^2$ at
90\% confidence level is obtained for a WIMP mass of 40 GeV/c$^{2}$.  This
represents more than a factor of two improvement on the best available
limits at this and higher masses. These improved cross-section limits
provide more stringent constraints on the effective WIMP-proton and
WIMP-neutron couplings.
\end{abstract}
\maketitle

The existence of dark matter (DM) in the Universe has been established
by numerous pieces of astronomical and cosmological evidence. These
range from the dynamics, gravitational lensing, and clustering of
galaxies to the necessity of DM to explain the power spectrum of the
cosmic microwave background and the formation of cosmological
structures. However, the particle nature of DM still remains
elusive. Weakly interacting massive particles (WIMPs), a class of
hypothetical particles predicted by many extensions of the Standard
Model of particle physics, are promising candidates for DM. Generic
WIMP production and annihilation rates in the early universe would
lead to a freeze-out WIMP density which could explain the observed DM
relic density (the so-called ``WIMP
miracle''~\cite{Jungman:1995df}). The detection of WIMP signals is the
goal of many past, ongoing and future experiments, including direct
detection experiments, indirect detection experiments, and experiments
at colliders.

The PandaX project consists of a series of xenon-based experiments,
located at the China JinPing underground Laboratory (CJPL). The first
two experiments, PandaX-I and PandaX-II, use xenon as a target to search
for WIMPs. The third experiment PandaX-III~\cite{Chen:2016qcd}, which
is being prepared, will search for neutrinoless double beta decay of
$^{136}$Xe. PandaX-I, with a 120-kg xenon target, was completed in
2014. PandaX-II, with a half-ton xenon target, has been running since
the end of 2015. Both the PandaX-I and PandaX-II experiments use a
dual-phase xenon time projection chamber technique. With this
technique, both the prompt scintillation photons (S1) produced in
liquid xenon and the delayed electroluminescence photons (S2) produced
in gaseous xenon for each physical event can be measured. This leads
to powerful background suppression and signal-background
discrimination. More detailed descriptions of the PandaX-I and
PandaX-II experiments can be found in
Refs.~\cite{Xiao:2014xyn,Xiao:2015psa,Tan:2016diz,Tan:2016zwf}.
 
The PandaX-II collaboration has recently reported WIMP search
results~\cite{Tan:2016zwf} using the first 98.7 days of data.  This
data set corresponds to a total exposure of 3.3$\times10^4$
kg-days. No excess of events was observed above the background, and
WIMP-nucleon cross-section upper limits were set assuming a 
spin-independent (SI) WIMP-nucleon interaction. The best upper limit
of $2.5\times10^{-46}$~cm$^2$ for a WIMP mass of 40 GeV/c$^{2}$ was
obtained. In this paper we consider an axial-vector, spin-dependent
(SD) interaction, which is well motivated if WIMPs have spin. An
example of this would be the lightest neutralino in the supersymmetric
theories, which offers one of the most promising DM
explanations. Xenon-based experiments, such as PandaX, XENON and LUX,
are sensitive to this interaction because xenon contains a significant
fraction of isotopes with non-zero spin. LUX
experiment~\cite{Akerib:2016lao}, with a total exposure of
1.4$\times10^4$ kg-days, pushed down the SD WIMP-neutron and
WIMP-proton cross-section limits to 9.4$\times10^{-41}$~cm$^2$ and
2.9$\times10^{-39}$~cm$^2$, respectively, at a WIMP mass of 33
GeV/c$^{2}$. XENON100 experiment~\cite{Aprile:2016swn} recently
updated their SD results with a total exposure of 1.8$\times10^4$
kg-days, obtaining slightly less stringent limits.

We use the same data set and identical event reconstruction and
selections as in Ref.~\cite{Tan:2016zwf}. Compared to ~\cite{Tan:2016zwf},
the data and expected background after selections remain unchanged. Below we
describe the WIMP-nucleus recoil rate calculation for the SD
WIMP-nucleon interaction, which will be needed to calculate the S1 and
S2 signal distributions and the final WIMP-nucleon cross-section upper
limits.
 
The nuclear recoil energy due to a WIMP with mass $m$ scattering
elastically from a nucleus with mass $M$ is
$E=(\mu^2v^2/M)(1-\cos\theta)$, where $\mu$ is the reduced mass, $v$
is the speed of the WIMP relative to the nucleus, and $\theta$ is the
scattering angle in the center of mass frame.  The differential event
rate with respect to recoil energy, in units of
counts$/$keV$/$day$/$kg of xenon, can be written
as~\cite{Savage:2008er}
\begin{equation} \frac{dR}{dE} =
\frac{\sigma^A(q)}{2m\mu^2}\rho\eta(E,t),
\label{drde}
\end{equation} where $q=\sqrt{2ME}$ is the nuclear recoil momentum,
$\sigma^A(q)$ is the WIMP-nucleus cross section, $\rho$ is the local
WIMP density, and $\eta(E,t)$ is the mean inverse speed of the time-dependent 
WIMP velocity distribution relative to the detector. The most frequently used 
distribution for the WIMP speed relative to the Milky way halo is a Maxwellian
distribution with the most probable value at $v_{0}=220$ km/s, and which is
truncated at the galactic escape velocity $v_{\mathrm{esc}}=544$
km/s. The calculation of Eq.~\ref{drde} follows the procedure in 
Ref.~\cite{Savage:2008er}. 

To report results for SD interactions, a common practice is to
consider the two limiting cases in which the WIMPs couple only to
protons or to neutrons. This practice is also consistent with the fact
that, due to the cancellation between spins of nucleon pairs, for
odd-A nuclei, $\sigma^A(q)$ is dominated by either contributions from
the unpaired proton (odd Z) or neutron (even Z). The intermediate cases can be
treated by following the methods in Ref.~\cite{Giuliani:2004uk}. In the two limiting cases, the SD
WIMP-nucleus cross section can then be written as
\begin{equation}
\sigma^A_{p,n}(q) = \frac{4\pi\mu^2S_{p,n}(q)}{3(2J+1)\mu^2_{p,n}}\sigma_{p,n},
\label{eq:xs}
\end{equation}
where $\mu_{p,n}$ is the WIMP-proton or WIMP-neutron reduced mass,
$\sigma_{p,n}$ is the WIMP-proton or WIMP-neutron cross section and
$J$ is the total angular momentum of the nucleus. Due to the above
mentioned spin pairing effects, the main Xe isotopes sensitive to SD
interactions are $^{129}$Xe ($J$= 1/2) and $^{131}$Xe ($J$=3/2). The
corresponding abundance in natural xenon is 26.4\% and 21.2\%,
respectively.

In Eq.~\ref{eq:xs} $S_{p,n}(q)$ is the spin structure factor for proton-only or
neutron-only coupling, obtained from nuclear shell model
calculations. In this paper we use the most recent calculation by Klos
\textit{et al.}~\cite{Klos:2013rwa} based on chiral effective
field theory at the one-body level, including the leading
long-range two-body currents. With this calculation, the ground states
and the ordering of the excited states of $^{129}$Xe and $^{131}$Xe
are very well described. This calculation was also used in recent SD
results from LUX~\cite{Akerib:2016lao} and
XENON100~\cite{Aprile:2016swn}. Alternative calculations by Ressell
and Dean~\cite{Ressell:1997kx} and by Toivanen \textit{et
al.}~\cite{PhysRevC.79.044302} generally do not agree with each other
nor with that by Klos \textit{et al.}~\cite{Klos:2013rwa}. A comparison 
of these calculations can be found in Ref.~\cite{Aprile:2013doa}.

\begin{figure}[h] \centering
  \includegraphics[width=0.45\textwidth]{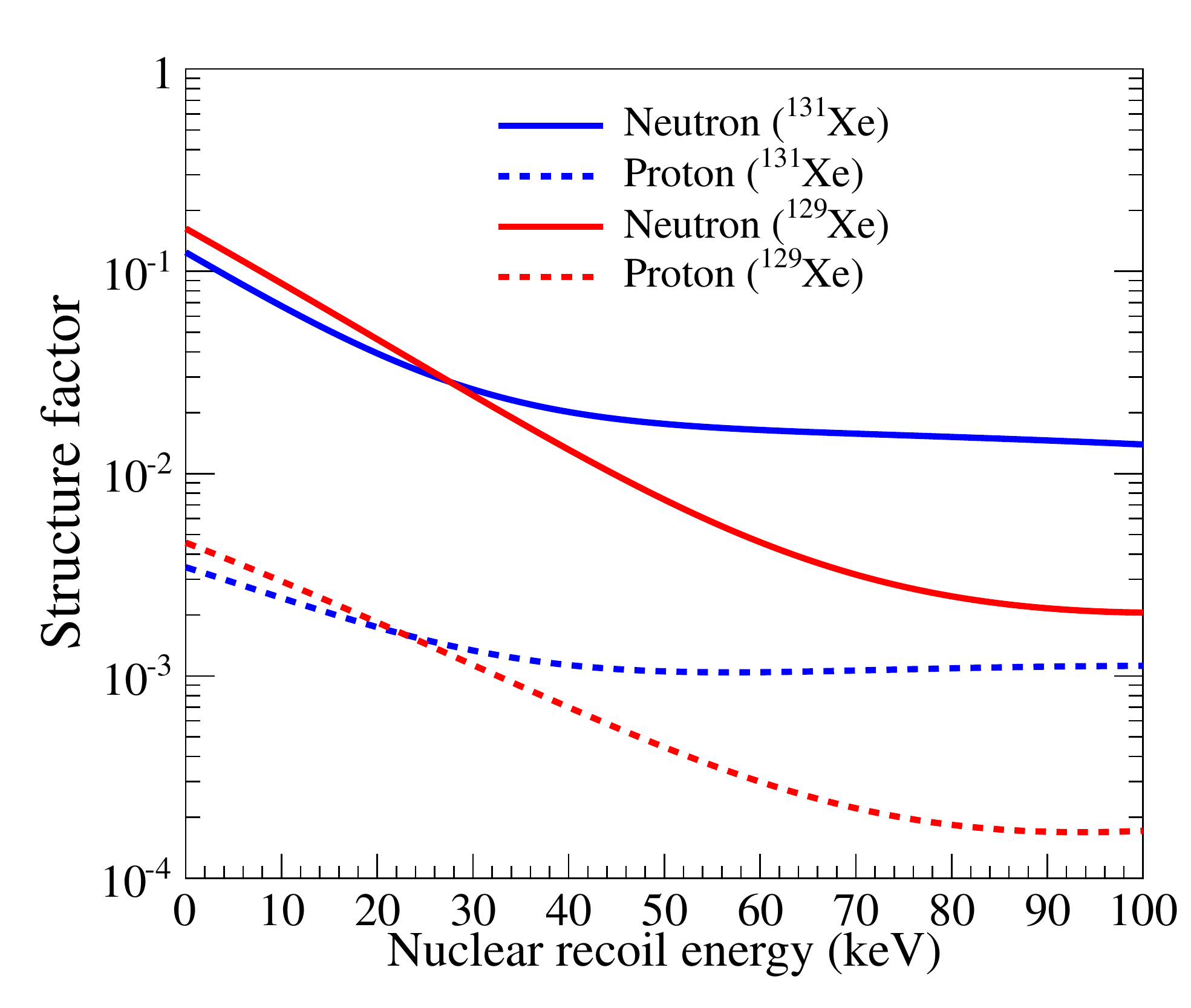}
  \caption{Structure factors as a function of nuclear recoil energy
for neutron-only (plain) and proton-only (dashed) couplings for
$^{129}$Xe (red) and $^{131}$Xe (blue), using the calculations from
Ref.~\cite{Klos:2013rwa}.}
  \label{fig:SA}
\end{figure}

For illustration, we compare structure factors using the calculation
from Ref.~\cite{Klos:2013rwa} as a function of nuclear recoil energy
for proton-only and neutron-only couplings. This is shown in Fig.~\ref{fig:SA}. For
both $^{129}$Xe and $^{131}$Xe, the neutron-only structure factor is
much larger than proton-only, since the total nuclear spin is
dominated by the unpaired neutron.  It is worth noting (as in
Ref.~\cite{Klos:2013rwa}) that ``neutron/proton-only'' is simply a
notation for convenience. When two-body currents are included,
neutrons can contribute to the proton-only coupling. This in fact
significantly enhances the proton-only while slightly reducing the
neutron-only structure factor.

\begin{figure}[h] \centering
  \includegraphics[width=0.45\textwidth]{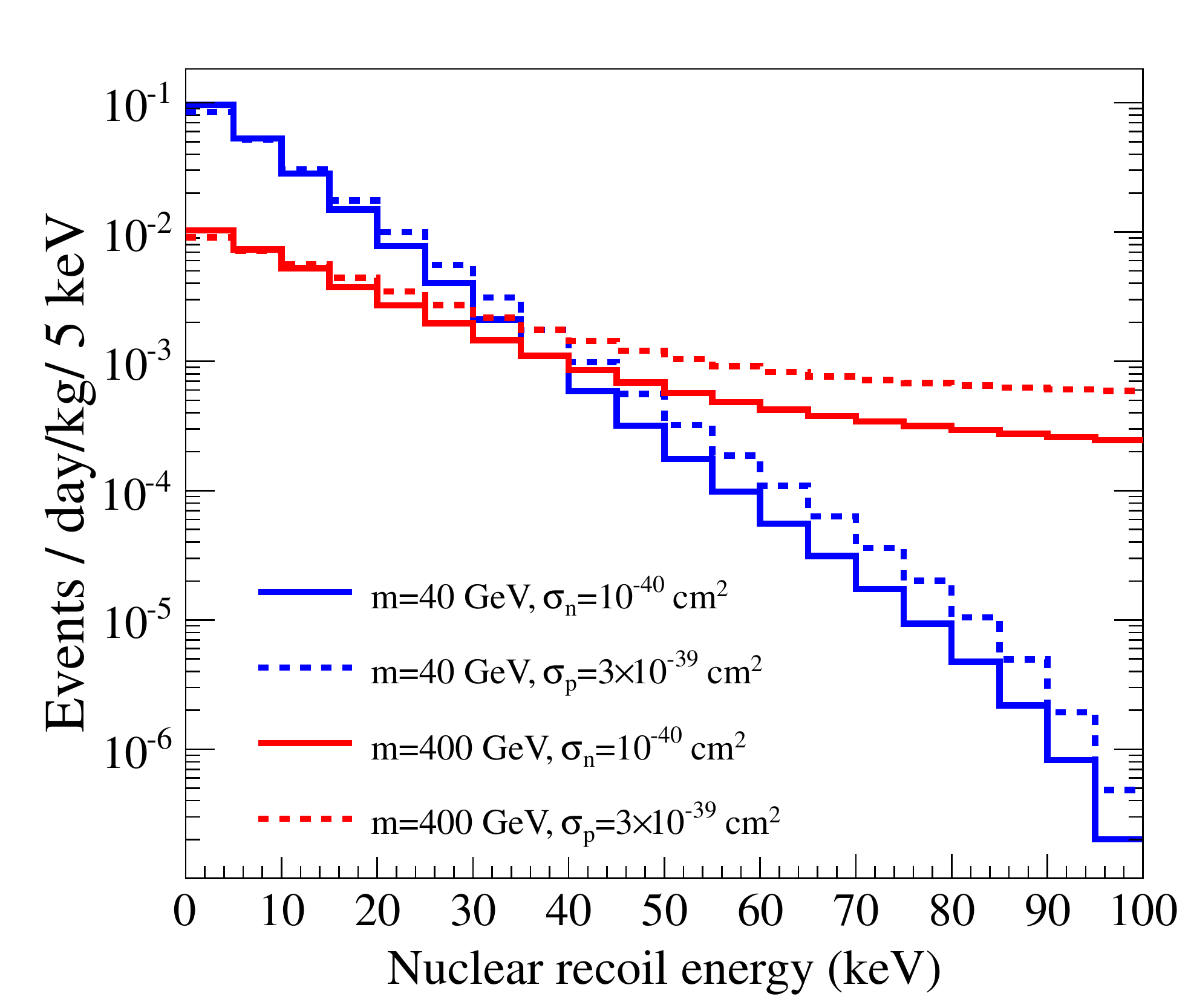}
  \caption{Nuclear recoil-energy distributions without detector
effects for two WIMP mass points 40 GeV/c$^{2}$ (blue) and 400
GeV/c$^{2}$ (red), for neutron-only (plain) and proton-only (dashed)
couplings. Here we use $dR/dE$ calculations from Ref.~\cite{Savage:2008er}
and structure factor calculations from Ref.~\cite{Klos:2013rwa}. The WIMP-neutron and WIMP-proton cross sections are assumed to be
$\sigma_{n}=10^{-40}$ cm$^{2}$ and $\sigma_{p}=3\times10^{-39}$
cm$^{2}$, respectively, for visual clarity.}
  \label{fig:dRdE}
\end{figure}

Fig.~\ref{fig:dRdE} shows the calculated recoil-energy distributions
without detector effects for two WIMP masses 40 GeV/c$^{2}$ and
400 GeV/c$^{2}$, for neutron-only and proton-only couplings. Here the
WIMP-neutron and WIMP-proton cross sections are assumed to be
$\sigma_{n}=10^{-40}$ cm$^{2}$ and $\sigma_{p}=3\times10^{-39}$
cm$^{2}$, respectively. This allows the two cases be to compared directly. The
recoil-energy distribution for proton-only coupling is harder than
neutron-only, since the proton-only structure factor decreases more slowly
at high recoil energies compared to the neutron-only.
 
\begin{figure}[h] \centering
  \includegraphics[width=0.45\textwidth]{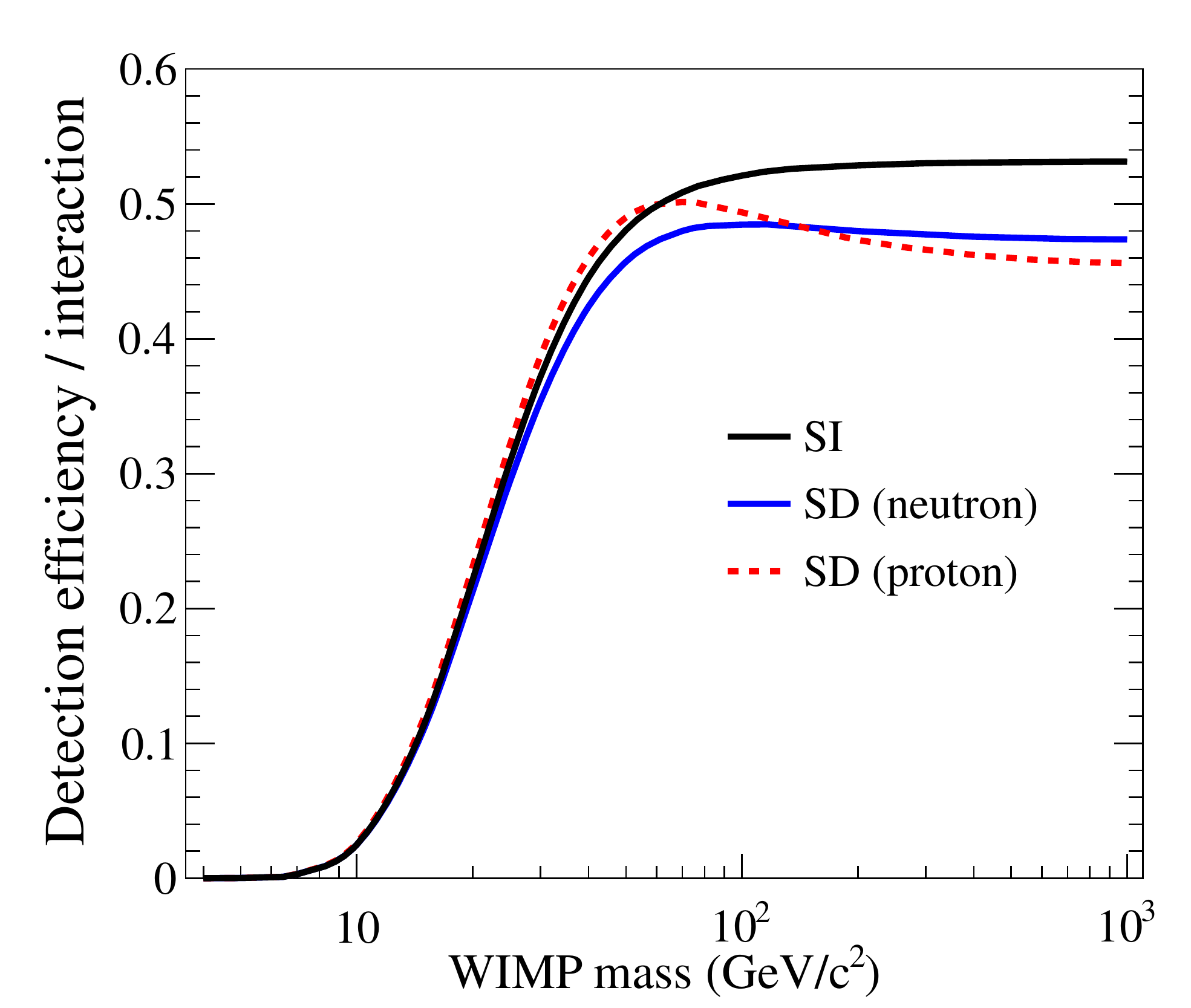}
  \caption{Detection efficiency per interaction as a function of the WIMP mass for neutron-only (blue) and proton-only (red) SD interactions. Efficiency for SI interaction (black) is also plotted for comparison.}
  \label{fig:eff}
\end{figure}

As in Ref.~\cite{Tan:2016zwf}, the data set is divided into 15
time bins to take into account the temporal change of detector
parameters and background rates. For each time bin, we simulate S1 and
S2 signal distributions from the obtained WIMP nuclear recoil-energy
distributions using the NEST model~\cite{Lenardo:2014cva}.  We apply
the same S1 and S2 selections as in Ref.~\cite{Tan:2016zwf}, requiring
an S1 in the range of 3 photoelectrons (PE) and 45 PE and an S2 in the
range of 100 PE (uncorrected) and 10000 PE. Fig.~\ref{fig:eff} shows
the final detection efficiency per WIMP-xenon interaction (weighted
average of 15 time bins) as a function of the WIMP mass for the 
neutron-only and proton-only SD interactions. The efficiency for SI
interaction is also included for comparison. Here, all measured
efficiencies for data quality, S1 and S2 selections, as well as the
additional boosted-decision-tree selection for suppressing accidental
background have been taken into account. For SD interaction, the
efficiency increases from approximately 2\% ($m=10$ GeV/c$^{2}$) to approximately 45\%
at high masses. The difference between proton-only and neutron-only
couplings is due to the difference of the recoil-energy distributions
(Fig.~\ref{fig:dRdE}) and the dependence of the efficiency on recoil
energy (Fig. 2 in Ref.~\cite{Tan:2016zwf}).

\begin{figure}[h] \centering
  \includegraphics[width=0.45\textwidth]{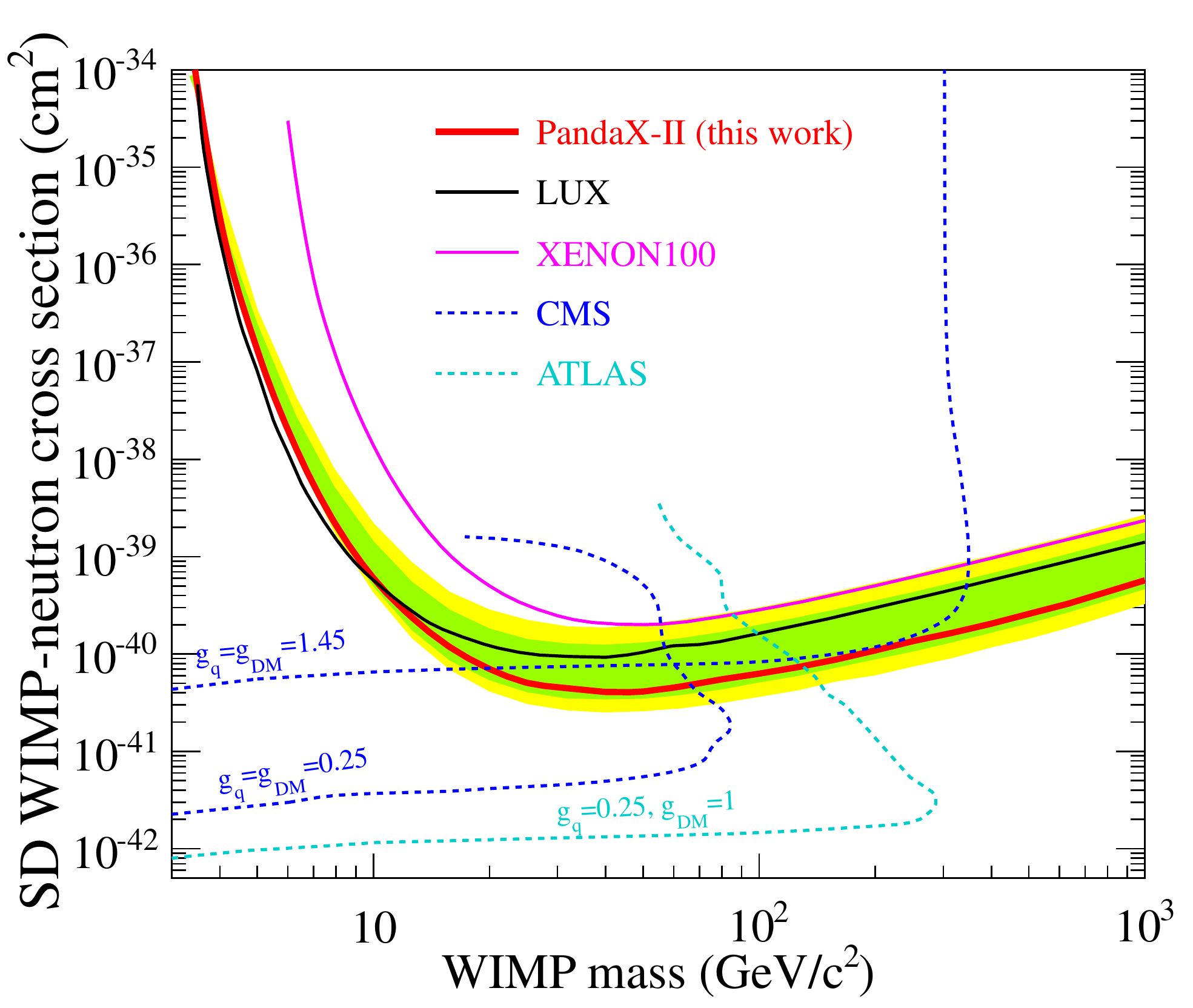}
  \includegraphics[width=0.45\textwidth]{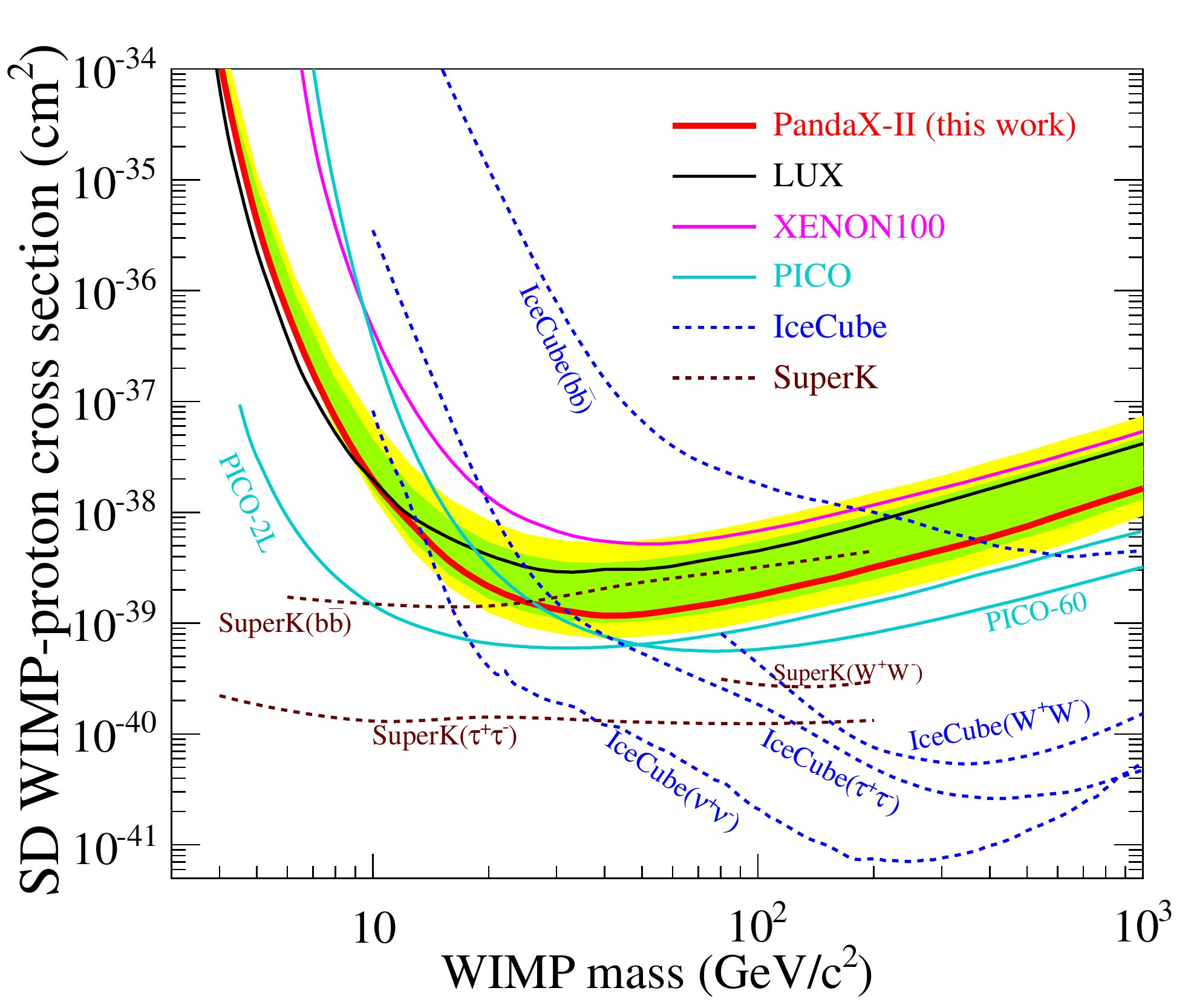}
\caption{PandaX-II 90\% CL upper limits for the SD WIMP-neutron (top)
and WIMP-proton (bottom) cross sections. Selected recent world results
are plotted for comparison: LUX~\cite{Akerib:2016lao},
XENON100~\cite{Aprile:2016swn}, CMS
mono-jet~\cite{Khachatryan:2014rra,Malik:2014ggr}, ATLAS
mono-jet~\cite{Aaboud:2016tnv},
PICO-2L~\cite{Amole:2016pye},PICO-60~\cite{Amole:2015pla},IceCube~\cite{Aartsen:2016exj}
and Super-K~\cite{PhysRevLett.114.141301}. The 1 and 2-$\sigma$
sensitivity bands are shown in green and yellow, respectively.  }
\label{fig:lim}
\end{figure}

The upper limits for the SD WIMP-nucleon cross sections are calculated
with the same procedure as in Ref.~\cite{Tan:2016zwf}. Test statistics
based on profile likelihood ratio ~\cite{Cowan:2010js,Aprile:2011hx}
were constructed over grids of WIMP mass and cross section. Then, the
90\% confidence level (CL) upper limits of cross sections were
calculated using the CL$_{\mathrm{s}}$ approach~\cite{CLS1,CLS2}. The
results are shown in Fig.~\ref{fig:lim}, with recent results from
other experiments overlaid. The upper limits presented lie within the
$\pm$1$\sigma$ sensitivity band. The lowest cross-section limit
obtained is 4.1$\times$10$^{-41}$~cm$^2$
(1.2$\times$10$^{-39}$~cm$^2$) for WIMP-neutron (WIMP-proton) elastic
scattering at a WIMP mass of 40 GeV/c$^{2}$.  For neutron-only
coupling, the lowest exclusion limits for WIMP above 10
GeV/c$^{2}$ in direct detection experiments are obtained.  Under model assumptions,
results from DM searches at colliders can also be interpreted as the
WIMP-nucleon cross-section limits. For example, mono-jet search
results from CMS~\cite{Khachatryan:2014rra} and
ATLAS~\cite{Aaboud:2016tnv} have been interpreted in the framework of
the so-called ``simplified'' DM
model~\cite{Buchmueller:2014yoa,Malik:2014ggr,Abercrombie:2015wmb}
which includes four parameters: the DM mass, the mediator mass, the
coupling of the mediator to DM particles ($g_{\mathrm{DM}}$) and the
coupling of the mediator to quarks ($g_{q}$). Four coupling scenarios
$g_{q}=g_{\mathrm{DM}}=0.25,0.5,1.0$ and 1.45 have been considered in
Ref.~\cite{Malik:2014ggr} for interpreting the CMS results. The ATLAS
collaboration reported the limits for the couplings $g_{q}=0.25$ and
$g_{\mathrm{DM}}=1$. In Fig.~\ref{fig:lim} we include the limits
obtained from the smallest and the largest couplings from CMS and the
limits from ATLAS. These limits are particularly strong for low mass
WIMPs, but one should note the strong model dependence. Our SD
WIMP-proton cross-section limits are much weaker than the WIMP-neutron
ones due to the even number of protons and the unpaired neutron in
$^{129}$Xe and $^{131}$Xe nuclei. PICO
experiments~\cite{Amole:2016pye,Amole:2015pla}, on the other hand,
utilizing $^{19}$F nuclei that contains unpaired protons, produced
so far the most stringent constraints on the SD WIMP-proton cross
sections in all direct search experiments. Indirect search experiments, 
IceCube~\cite{Aartsen:2016exj} and Super-K~\cite{PhysRevLett.114.141301}, 
can produce more stringent limits, depending on WIMP masses and annihilation channels.

\begin{figure}[h] \centering
  \includegraphics[width=0.45\textwidth]{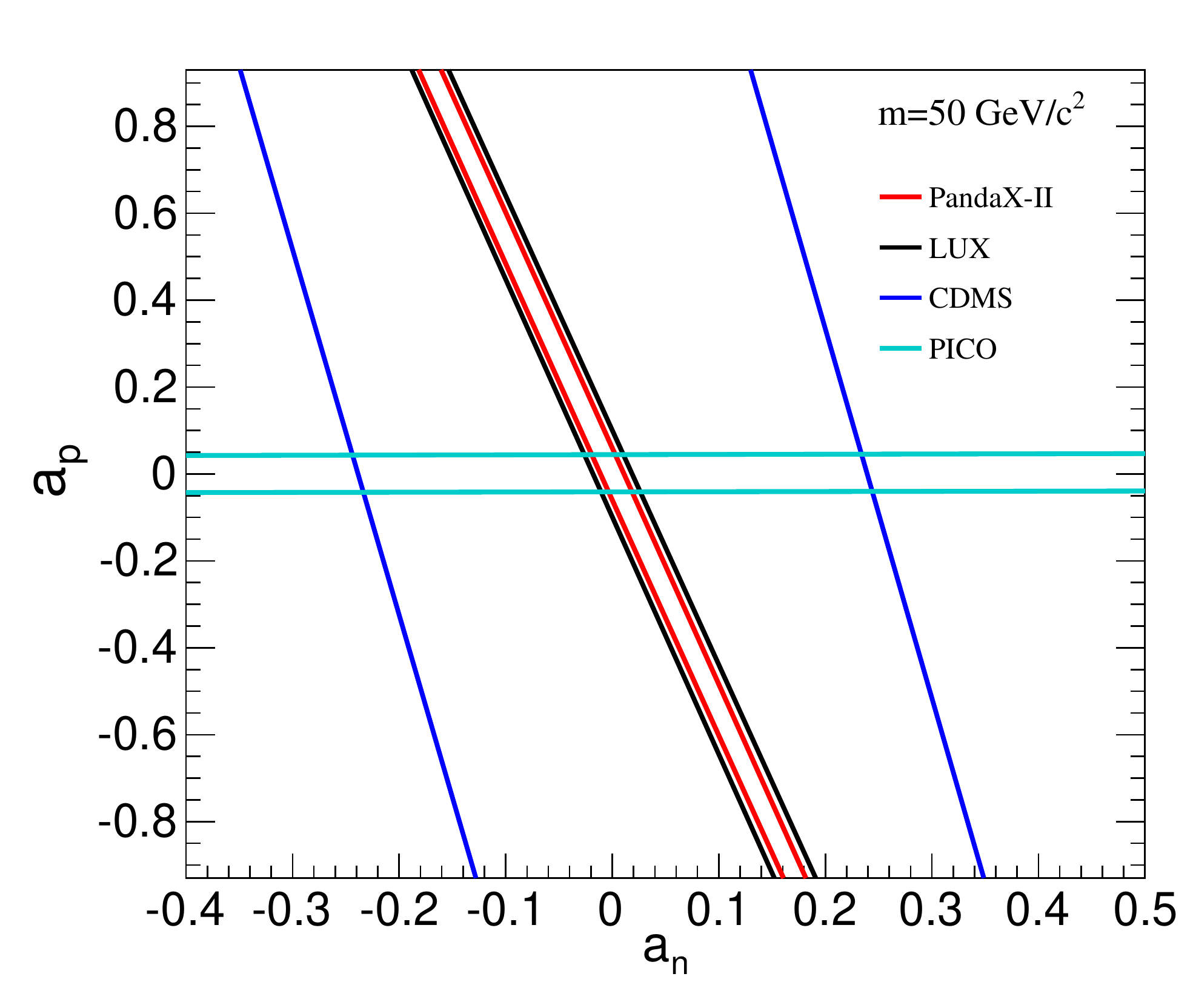}
  \includegraphics[width=0.45\textwidth]{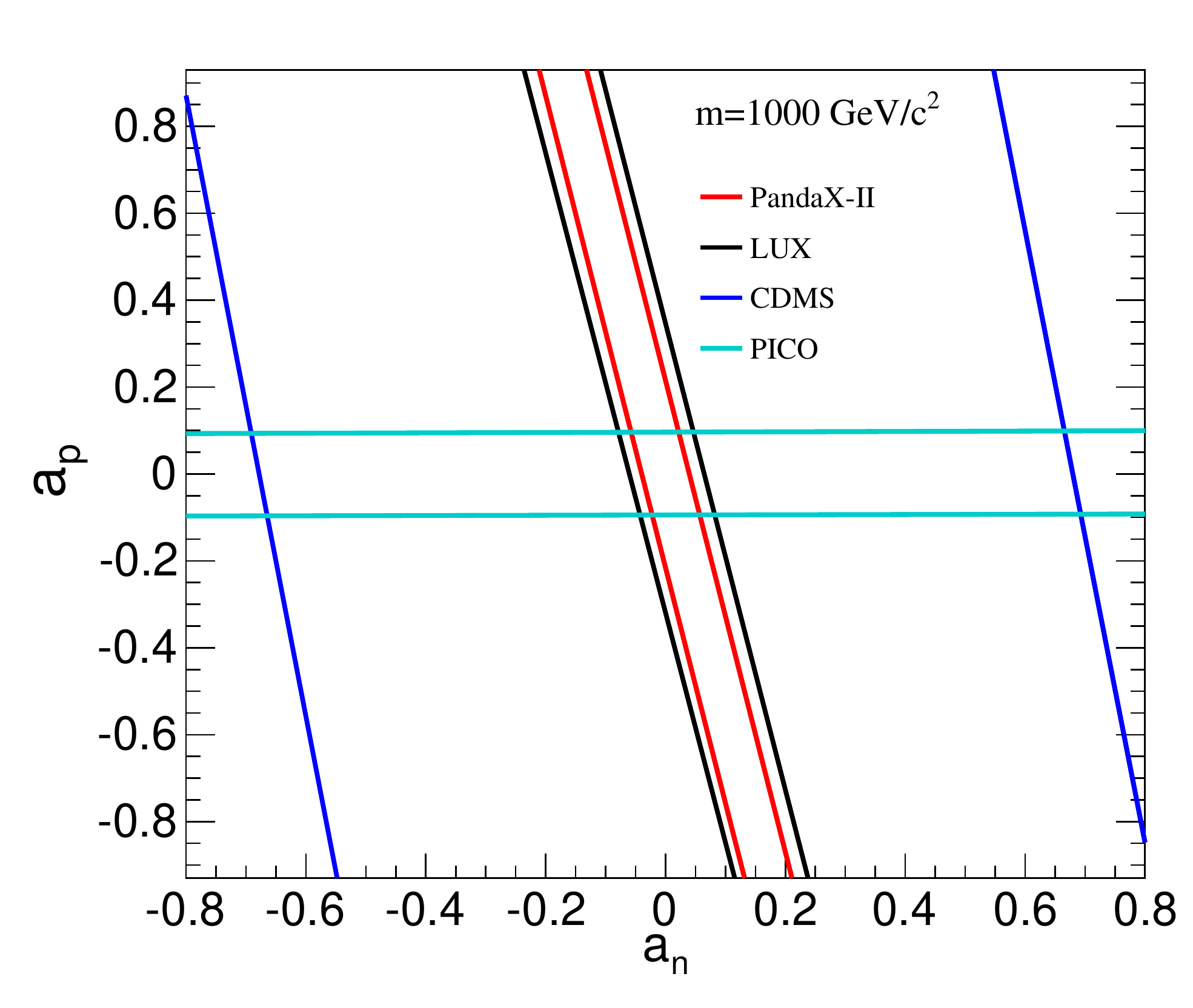}
\caption{PandaX-II constraints on the effective WIMP-proton and
WIMP-neutron couplings, $a_p$ and $a_n$, for two WIMP masses (50 and
1000 GeV/c$^2$).  Also shown are results from LUX, PICO and CDMS
experiments (all from Ref.~\cite{Akerib:2016lao}). }
\label{fig:2dcoupling}
\end{figure}

The WIMP-neutron and WIMP-proton cross-section upper limits can be
used to constrain the effective WIMP couplings to neutrons and
protons, $a_{n}$ and $a_{p}$, simultaneously. For a given WIMP mass,
the allowed region in the $a_p$-$a_n$ plane is derived
from~\cite{Tovey:2000mm,Giuliani:2005bd}
\begin{equation}
\sum_{A}(\frac{a_{p}}{\sqrt{\sigma^{\mathrm{lim}(A)}_{p}}}\pm\frac{a_{n}}{\sqrt{\sigma^{\mathrm{lim}(A)}_{n}}})^{2}
\leq \frac{\pi}{24G^{2}_{F}\mu^{2}_{p}},
\end{equation} where ${\sigma^{\mathrm{lim}(A)}_{p,n}}$ are the upper
limits of the WIMP-proton and WIMP-neutron cross sections for the
isotope with mass number A. Fig~\ref{fig:2dcoupling} shows the allowed
region in the $a_p$-$a_n$ plane, together with results from LUX, PICO, and 
CDMS experiments (all calculated in Ref.~\cite{Akerib:2016lao}) for two WIMP masses (50 and 1000
GeV/c$^{2}$). This shows our improvement over previous results, as well as the 
complementarity between experiments with different detection mediums.

In conclusion, the 90\% CL upper limits of the SD
WIMP-nucleon cross section using recently released data of the
PandaX-II experiment with a total exposure of
3.3$\times$10$^4$~kg-days have been presented. For WIMPs with masses above 10 GeV/c$^{2}$, 
the most stringent upper limits to date on the SD WIMP-neutron
cross sections in all direct DM search experiments are set, with a lowest
excluded value of 4.1$\times$10$^{-41}$~cm$^2$ at a WIMP mass of 40
GeV/c$^{2}$. This result is complementary to the results obtained from
WIMP searches performed at the LHC, which can produce strong limits
particularly for low mass WIMPs, depending on the models and
assumptions. For high mass WIMPs, the constraints on the effective WIMP-proton
and WIMP-neutron couplings have also been improved over previous 
results from direct DM search experiments. These constraints are complementary to experiments (such as PICO), which are more sensitive to WIMP-proton than WIMP-neutron coupling.  

\begin{acknowledgments}
  This project has been supported by a 985-III grant from Shanghai
Jiao Tong University, grants from the National Science Foundation of China
(Nos. 11435008, 11455001, 11505112 and 11525522), and a grant from the
Office of Science and Technology in Shanghai Municipal Government
(No. 11DZ2260700). This work is supported in part by the Chinese
Academy of Sciences Center for Excellence in Particle Physics (CCEPP),
the Key Laboratory for Particle Physics, Astrophysics and Cosmology,
Ministry of Education, and Shanghai Key Laboratory for Particle
Physics and Cosmology (SKLPPC).  The project is also sponsored by
Shandong University, Peking University, and the University of
Maryland. We thank C. Hall for helping with the specifics of tritium
calibration. We thank P. Klos \textit{et al.} for providing their
structure factor calculations.  We also thank the CJPL administration
and the Yalong River Hydropower Development Company Ltd for
indispensable logistics and other support. Finally, we thank the Hong
Kong Hongwen Foundation for financial support.

\end{acknowledgments}

\bibliographystyle{apsrev4-1}
\bibliography{refs}

\end{document}